# NEW INSIGHTS ON THE OPTIMALITY OF PARAMETERIZED WIENER FILTERS FOR SPEECH ENHANCEMENT APPLICATIONS

Rafael Attili Chiea

Department of Electrical and Electronic Engineering, Federal University of Santa Catarina,

Florianópolis-SC, 88040-900, Brazil. Tel.: +55 48 3721-9506, E-mail: rafaelchiea@gmail.com.

Márcio Holsbach Costa[*]

Department of Electrical and Electronic Engineering, Federal University of Santa Catarina,

Florianópolis-SC, 88040-900, Brazil. Tel.: +55 48 3721-9506, E-mail: costa@eel.ufsc.br.

Guillaume Barrault

Wavetech ST, Florianópolis-SC, 88030-035, Brazil. Tel.: +55 48 3025-5858, E-mail:

guillaume@wavetech-st.com.

[*] Corresponding author

# ABSTRACT

This work presents a unified framework for defining a family of noise reduction techniques for speech enhancement applications. The proposed approach provides a unique theoretical foundation for some widely-applied soft and hard time-frequency masks, which encompasses the well-known Wiener filter and the heuristically-designed Binary mask. These techniques can now be considered as optimal solutions of the same minimization problem. The proposed cost function is defined by two design parameters that not only establish a desired trade-off between noise reduction and speech distortion, but also provide an insightful relationship with the mask morphology. Such characteristic may be useful for applications that require online adaptation of the suppression function according to variations of the acoustic scenario. Simulation examples indicate that the derived conformable suppression mask has approximately the same quality and intelligibility performance capability of the classical heuristically-defined parametric Wiener filter. The proposed approach may be of special interest for real-time embedded speech enhancement applications such as hearing aids and cochlear implants.

**KEYWORDS:** Noise reduction; speech enhancement; time-frequency mask; hearing aids; cochlear implant.

# 1. INTRODUCTION

Speech enhancement methods have been a subject of great interest by the signal processing community for many years. They are a fundamental part of a wide variety of applications, ranging from automatic speech recognition systems to speech coding and assistive hearing devices. Their main objectives are to increase intelligibility during communication, and to improve speech quality and acoustical comfort, avoiding fatigue due to noisy speech [1]. These techniques are usually grouped into single or multichannel approaches (according to the number of input microphones), and are conceived not only to reduce noise levels but also to emphasize some specific speech characteristics. When designed in the light of signal distortion and noise suppression requirements they are usually referred to as noise reduction methods.

Although it has been demonstrated that multichannel speech enhancement methods may improve speech understanding and localization of acoustical sources, there is still great interest on single-channel processing due to cost, size and power consumption constraints required by many embedded applications.

Noise reduction techniques are generally applied in the time-frequency framework, in which the decomposition of the input (noisy) signal into multiple frequency bands is processed by a point wise multiplication of an attenuation factor (at each frame and bin), also referred to as gain. The set of these attenuation factors, which is associated to a certain suppression rule, is called mask. After processing, the estimated (clean) speech is obtained by transforming back the signal to the time domain.

Two main approaches are commonly employed for designing time-frequency masks: the heuristic and the formal approaches. In the latter, the minimization of a desired cost function, usually associated to a trade-off between noise reduction and speech distortion, is performed. The formal approach has the advantage of not only being associated to certain logic, but also to a theoretical justification. The most popular time-frequency mask techniques are, undoubtedly, the Wiener filter (WF) and the Binary mask (BM) [1] [2] [3] [4].

The Binary mask was proposed by Cooke et al. in 2001 in the context of complex auditory scene analysis (CASA) applications [5]. The idea is to suppress noise-dominant time-frequency units, keeping information in which the target signal power is dominant over noise [3] [6]. Such approach counts on the ability of the human hearing to deal with missing data to reconstitute audio cognition.

In contrast to the heuristically-defined hard-decision approach related to the BM, the Wiener filter is the optimal solution that minimizes the mean square error between the desired and estimated signals. It is characterized by a soft-decision mask that provides continuous gains from 0 to 1. It has been shown that the WF provides improved speech quality as compared to BM, and, for hearing impaired listeners it may outperform the BM in terms of speech intelligibility, for both ideal and perturbed gain estimates [3] [7]. Despite being mathematically optimal, the Wiener filter is not always the best solution in terms of speech perception, since it may introduce undesirable musical noise [8]. In order to improve its psychoacoustic performance, heuristically designed versions of the Wiener filter have been proposed [9] [10] in which the inclusion of extra parameters allows more flexibility on the morphology of the mask [11] [12].

The Parametric Wiener (PW) filter, which was introduced in the context of speech enhancement applications in [9], is a generalization of the classical Wiener filter. Two extra parameters were included to allow further conformability, making it a benchmark for performance analysis of noise reduction techniques [10]. Despite some efforts in trying to provide a deeper interpretation for particular situations [13] [14], there is still no theoretical support for optimality of its general configuration. In this way, since its proposition in 1979, it is still considered an *ad-hoc* technique [10] [14]. Thereby, the setting of its parameters is performed empirically [12] [15] [16] or based on *a posteriori* rules [17] [18] [19].

In this work, we propose a new cost function, based on a trade-off between speech distortion and noise reduction, for designing time-frequency masks in speech enhancement applications. It is shown that the optimal solution for this formal framework results in a previous heuristically-defined version of the parametric Wiener filter [14]. Here, it is demonstrated that it generates a whole family of suppression rules, which comprises not only the classical BM and WF but also other well-established approaches, allowing not only a manifold performance investigation, but also an insightful relationship between the design parameters and the shape of the suppression function. Speech quality and intelligibility objective measures are applied to illustrate the applicability and performance of the resulting conformable mask (CM).

The novel contributions of this work are: 1) The proposition of a general two-parameter cost function for establishing a trade-off between speech distortion and noise reduction in speech enhancements applications; 2) The theoretical derivation of the optimal solution for the proposed cost function, resulting in a family of suppression rules defined according to the choice of the parameter setting; 3) The demonstration that both the Wiener filter and the Binary mask are particular solutions of the same optimization problem; 4) The demonstration that the Binary mask is the optimal solution of a well-defined minimization problem; and 5) To provide an interpretation about the relationship between the parameters and the morphology of the proposed suppression function. Simulation examples indicate that the derived suppression mask has approximately the same performance capability of the classical heuristically-defined parametric Wiener filter [9].

The remainder of this paper is structured as follows: Section 2 contains the problem definition and presents some suppression masks widely applied in the literature. The proposed cost function and its optimal solution are presented in Section 3. Experimental methods and simulation results are presented in Section 4, while discussion is presented in Section 5. Finally, concluding remarks are presented in Section 6.

## 2. PROBLEM DEFINITION

Let us consider that noisy speech is defined as $y(n) = x(n)+v(n)$, in which $x(n)$ is the desired speech signal, and $v(n)$ is the additive noise. Both $x(n)$ and $v(n)$ are considered not individually observable and uncorrelated to each other. Taking the $N$-bin Short-Time-Fourier-Transform (STFT) representation for a finite time-window of $y(n)$ results in:

$$Y(k,\lambda) = X(k,\lambda) + V(k,\lambda) , \qquad (1)$$

in which $k$ and $\lambda$ are, respectively, the frequency-band and the time-frame indexes; and $Y(k,\lambda)$, $X(k,\lambda)$, and $V(k,\lambda)$ are respectively the STFTs of $y(n)$, $x(n)$, and $v(n)$.

The time-frequency speech-enhancement approach consists of (at each time-frame $\lambda$) multiplying the noisy signal $Y(k,\lambda)$ by a gain mask $M(k,\lambda)$, for generating an estimate to the target speech $\hat{X}(k,\lambda)$ (at each $k$-bin) in a way that

$$\hat{X}(k,\lambda) = M(k,\lambda) Y(k,\lambda) . \qquad (2)$$

The estimated speech signal is reconstructed to the time domain by an overlap-and-add strategy.

### 2.1. Classical Suppression Masks

There are different approaches for defining the gain mask $M(k,\lambda)$. They may be defined according to a chosen objective criterion or even by a heuristic approach. In general, they are based on functions of signal-to-noise-ratio (SNR) estimates of the noisy signal. The most simple suppression function is the Binary mask, defined as [4]:

$$M(k,\lambda) = B(k,\lambda) = \begin{cases} 1 & : \xi(k,\lambda) \geq \mu_0 \\ 0 & : \xi(k,\lambda) < \mu_0 \end{cases}, \qquad (3)$$

in which $\xi(k,\lambda) = \sigma_X^2(k,\lambda)/\sigma_V^2(k,\lambda)$ is the *a priori* SNR associated to each frequency-band $k$ and time-frame $\lambda$; $\sigma_X^2(k,\lambda) = E\{|X(k,\lambda)|^2\}$ and $\sigma_V^2(k,\lambda) = E\{|V(k,\lambda)|^2\}$ are the spectral density functions of the clean speech and noise; $E\{\cdot\}$ is the expected value; $|\cdot|$ is the absolute value; and $\mu_0$ is the decision threshold, usually set to 0 dB. Due to the discontinuity on their (binary) gains, it is classified as a hard mask.

The soft masks are those characterized by suppression functions with smooth transitions between extreme values. The most popular soft mask is the Wiener filter [1], whose gain function is given by

$$M(k,\lambda) = W(k,\lambda) = \frac{\xi(k,\lambda)}{\xi(k,\lambda) + 1}. \qquad (4)$$

The Wiener filter is the optimal solution that minimizes the mean square error (MSE) between the STFT of the estimated and desired signals: MSE $= E\{|\hat{X}(k,\lambda) - X(k,\lambda)|^2\}$, assuming stationary signals.

Other examples of well-known soft masks are the constrained Wiener filter [1]

$$M(k,\lambda) = W_c(k,\lambda) = \frac{\sqrt{\xi(k,\lambda)}}{\sqrt{\xi(k,\lambda)}+1}, \qquad (5)$$

and the parametric Wiener filter [9]:

$$M(k,\lambda) = W_p(k,\lambda) = \left(\frac{\xi(k,\lambda)}{\xi(k,\lambda)+\eta}\right)^\beta. \qquad (6)$$

The parametric Wiener filter is a heuristically defined soft-weighting mask that is used as a benchmark performance for comparison of time-frequency masks. For the particular case of $\beta = 1$ and $\eta = 1$ it becomes the Wiener filter [1].

All presented masks require estimates of the *a priori* SNR that are commonly obtained by applying the decision-directed method or its variations [20] [21].

## 3. PROPOSED CONFORMABLE MASK

In this section, we propose a new optimization framework for deriving time-frequency masks. It is defined by a cost function that is based on an arbitrary trade-off between speech distortion and noise reduction. Its minimization provides a family of suppression functions that keeps intuitive relationships between the mask shape and the design parameters.

### 3.1. Cost Function

Assuming a linear estimator of the target speech signal, then $\hat{X}(k,\lambda) = H(k,\lambda)Y(k,\lambda)$, in which $H(k,\lambda) \in \mathbb{C}$. The estimation error is given by:

$$\begin{aligned}\varepsilon &= \hat{X}(k,\lambda) - X(k,\lambda) \\ &= H(k,\lambda)Y(k,\lambda) - X(k,\lambda)\end{aligned}. \qquad (7)$$

Using (1) in (7) results in [1]

$$\begin{aligned}\varepsilon &= H(k,\lambda)[X(k,\lambda)+V(k,\lambda)] - X(k,\lambda) \\ &= \underbrace{[H(k,\lambda)-1]X(k,\lambda)}_{\varepsilon_X(k,\lambda)} + \underbrace{H(k,\lambda)V(k,\lambda)}_{\varepsilon_V(k,\lambda)}\end{aligned}, \qquad (8)$$

in which $\varepsilon_X(k,\lambda)$ and $\varepsilon_V(k,\lambda)$ are, respectively the speech and noise distortions in the frequency domain [1] [22]. Assuming speech and noise are stationary signals in a given time-window, the power spectral densities of both $\varepsilon_X(k,\lambda)$ and $\varepsilon_V(k,\lambda)$ are defined as [1]

$$d_X(k,\lambda) = E\{\varepsilon_X^2(k,\lambda)\} = |H(k,\lambda)-1|^2 \sigma_X^2(k,\lambda) , \tag{9}$$

$$d_V(k,\lambda) = E\{\varepsilon_V^2(k,\lambda)\} = |H(k,\lambda)|^2 \sigma_V^2(k,\lambda) . \tag{10}$$

From (9) and (10), the proposed conformable cost function is defined as

$$J(k,\lambda) = [d_X(k,\lambda)]^\alpha + \rho[d_D(k,\lambda)]^\alpha , \tag{11}$$

in which $\rho$ is the relative weighting factor between speech distortion ($d_X$) and noise reduction ($d_V$); and $\alpha$ is the cost function steepness.

### 3.2. Optimal Solution

Equations (9) and (10) are both analytic convex functions of $H(k,\lambda)$. Thus, considering $\rho \in \mathbb{R}_{>0}$ and $\alpha \in (\frac{1}{2}, \infty)$, equation (11) is also analytic and convex with respect to $H(k,\lambda)$ [23]. This implies on the existence of only one $H(k,\lambda)$ that globally minimizes $J(k,\lambda)$. The optimal solution for $J(k,\lambda)$ can be obtained by differentiating (11) with respect to $H^*(k,\lambda)$ and equating it to zero:

$$\frac{\partial J(k,\lambda)}{\partial H^*(k,\lambda)} = 0 , \tag{12}$$

where $(\cdot)^*$ is the complex conjugate. Using (9), (10), and (11) in (12) leads to

$$\frac{\partial}{\partial H^*(k,\lambda)} \left\{ [H(k,\lambda)-1]^{2\alpha} \sigma_X^2(k,\lambda)^\alpha + \rho H(k,\lambda)^{2\alpha} \sigma_V^2(k,\lambda)^\alpha \right\} = 0 . \tag{13}$$

Differentiating (13) results in

$$\alpha \frac{\left[(H(k,\lambda)-1)(H(k,\lambda)-1)^*\right]^\alpha}{(H(k,\lambda)-1)^*} \sigma_X^2(k,\lambda)^\alpha + \alpha\rho \frac{\left[H(k,\lambda)H^*(k,\lambda)\right]^\alpha}{H^*(k,\lambda)} \sigma_V^2(k,\lambda)^\alpha = 0 , \tag{14}$$

and rearranging (14) leads to

$$\left[1 - \frac{1}{H(k,\lambda)}\right]^\alpha \left[1 - \frac{1}{H^*(k,\lambda)}\right]^{\alpha-1} = -\rho \frac{\sigma_V^2(k,\lambda)^\alpha}{\sigma_X^2(k,\lambda)^\alpha} . \tag{15}$$

Defining

$$Z(k,\lambda) \triangleq 1 - \frac{1}{H(k,\lambda)} , \qquad (16)$$

considering $\rho = \mu^\alpha$, and using $\xi(k,\lambda) = \sigma_X^2(k,\lambda)/\sigma_V^2(k,\lambda)$ in (15) results in

$$Z(k,\lambda)^\alpha \left(Z^*(k,\lambda)\right)^{\alpha-1} = -\left(\frac{\mu}{\xi(k,\lambda)}\right)^\alpha . \qquad (17)$$

Using the polar form of $Z(k,\lambda) = |Z(k,\lambda)|e^{j\phi_Z}$, and $e^{j2\pi p} = e^{j2\pi q} = e^{j2\pi l} = 1$ for $\{p, q, l\} \in \mathbb{Z}$ in (17) leads to

$$\frac{|Z(k,\lambda)|^\alpha e^{j\alpha\phi_Z} e^{j2\pi\alpha p} |Z(k,\lambda)|^\alpha e^{-j\alpha\phi_Z} e^{-j2\pi\alpha q}}{|Z(k,\lambda)| e^{-j\phi_Z}} = -\left|\frac{\mu}{\xi(k,\lambda)}\right|^\alpha e^{j2\pi\alpha l} . \qquad (18)$$

Rearranging (18) results in

$$|Z(k,\lambda)|^{2\alpha-1} e^{j\phi_Z} e^{j2\pi\alpha s} = -\left|\frac{\mu}{\xi(k,\lambda)}\right|^\alpha e^{j2\pi\alpha l} , \qquad (19)$$

in which $\{s = p-q, l\} \in \mathbb{Z}$. Comparing both modulus and phase of both sides of (19) results in:

$$e^{j\phi_Z} = -1 , \qquad (20)$$

and

$$|Z(k,\lambda)| = \left|\frac{\mu}{\xi(k,\lambda)}\right|^{\frac{\alpha}{2\alpha-1}} . \qquad (21)$$

Substituting (20) and (21) in $Z(k,\lambda) = |Z(k,\lambda)|e^{j\phi_Z}$, and finally in (16), leads to

$$H(k,\lambda) = \frac{\left|\frac{\xi(k,\lambda)}{\mu}\right|^{\frac{\alpha}{2\alpha-1}}}{1 + \left|\frac{\xi(k,\lambda)}{\mu}\right|^{\frac{\alpha}{2\alpha-1}}} . \qquad (22)$$

Since $\xi(k,\lambda)/\mu \in \mathbb{R}_{>0}$, it is possible to drop the modulus operator from (22) resulting in the optimal solution for (11):

$$H(k,\lambda) = \frac{\xi(k,\lambda)^\gamma}{\xi(k,\lambda)^\gamma + \mu^\gamma} . \qquad (23)$$

in which $\mu = \rho^{1/\alpha}$, and $\gamma = 1/(2-1/\alpha)$. Equation (23) was previously described in [14], without grounded theoretical foundations or experimental assessment, as a heuristic alternative to the (*ad hoc*) parametric Wiener filter.

### 3.3. Conformability Analysis

According to Section 3.2 the design parameter ranges are set to $\rho \in \mathbb{R}_{>0}$ and $\alpha \in (½, \infty)$, resulting in $\mu \in \mathbb{R}_{>0}$ and $\gamma \in (½, \infty)$. By adjusting these parameters the proposed conformable mask $H(k,\lambda)$ presented in (23) defines a family of sigmoidal masks, which include the classic WF mask ($\mu = \gamma = 1$) and the ideal BM ($\mu = 1$ and $\gamma \to \infty$), as shown in Table I.

Table I. Relationship among the proposed and some well-established time-frequency masks as a function of the design parameters $\mu$ and $\gamma$.

| $\gamma$ | $\mu$ | $H(k,\lambda)$ |
|---|---|---|
| $\gamma \to \infty$ | $\mu = \mu_o$ | $H(k,\lambda) \to$ BM, eq. (3) |
| $\gamma = 1$ | $\mu = 1$ | $H(k,\lambda) \equiv$ WF, eq. (4) |
| $\gamma = 1$ | $\mu = \mu_o$ | $H(k,\lambda) \to$ spectrum over-subtraction method [24] |
| $\gamma \to ½_+$ | $\mu = 0$ | $H(k,\lambda) \to$ constrained WF, eq. (5) |

Fig. 1 shows the conformability of the proposed mask for different sets of $\mu$ and $\gamma$ as a function of the SNR. It is clearly verified that $\gamma$ is directly related to the maximum derivative (sharpness or slope) of the resulting mask, while $\mu$ controls the bias (lateral displacement or transition threshold) of the suppression function. Considering the Root Mean Square Error (RMSE) between two noise suppression masks, defined by

$$RMSE = \sqrt{\int_{-\infty}^{\infty}[G_1(\xi) - G_2(\xi)]^2 d\xi} \ , \qquad (24)$$

(where $G_1$ and $G_2$ define two different noise suppression functions) the matching error between the proposed mask for $\mu = 1$ and $\gamma = 100$ and the ideal binary mask is 10% of the error between the WF and the BM, while for $\mu = 1$ and $\gamma = 1000$ this error drops approximately to 3%[1]. From the obtained results, it is possible to verify that there is an intuitive relationship between the shape of the suppression mask and their design parameters.

---

[1] The RMSE was calculated by approximating the integral presented in (24) via the trapezoidal method for $-60$ dB $\leq$ SNR $\leq 60$ dB in steps of 0.001 dB.

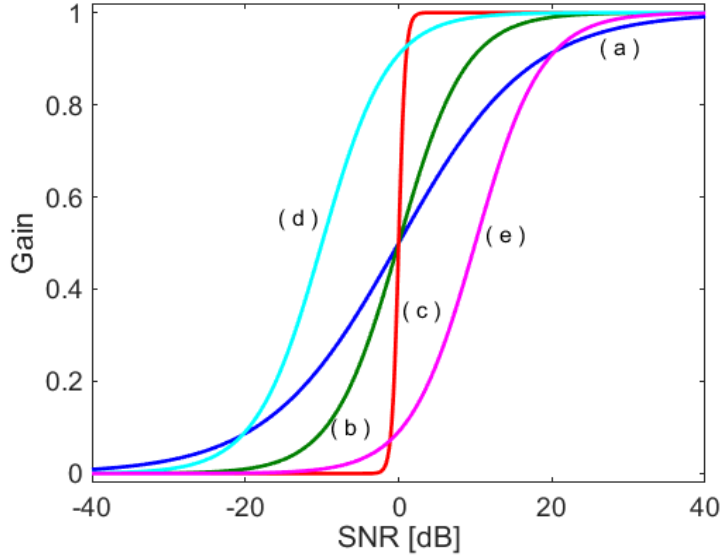

Fig. 1. Conformability of $H(k,\lambda)$ for different sets of $\mu$ and $\gamma$ as a function of the SNR. Parameter $\mu$ controls bias (lateral displacement), while parameter $\gamma$ is associated to the smoothness of the transition (slope). For $\mu = \gamma = 1$ $H(k,\lambda)$ turns to the classical Wiener filter mask, while for $\mu = 1$ and $\gamma \rightarrow \infty$ it tends to the ideal binary mask. (a) $\gamma = 0.51$, $\mu = 0$ dB (blue); (b) $\gamma = 1$, $\mu = 0$ dB (green); (c) $\gamma = 10$, $\mu = 0$ dB (red); (d) $\gamma = 1$, $\mu = -5$ dB (cyan); (e) $\gamma = 1$, $\mu = 5$ dB (magenta).

## 4. SIMULATION RESULTS

In this section, speech quality and intelligibility objective measures are applied for performance assessment capability of the proposed suppression function as compared to some classic speech enhancement techniques.

A set of 720 sentences from the balanced IEEE corpus [25] was artificially contaminated with either noise recorded inside a train wagon (Fig. 2a) or cafeteria babble noise (Fig. 2b) [1], for three levels of SNR: −10 dB, 0 dB, and 5 dB. This resulted in 4320 noisy signals. A total of 3720 of which were applied for assessment and performance comparison of the Wiener filter, the ideal Binary mask ($\mu_0 = 0$ dB), the proposed mask, and the parametric Wiener filter.

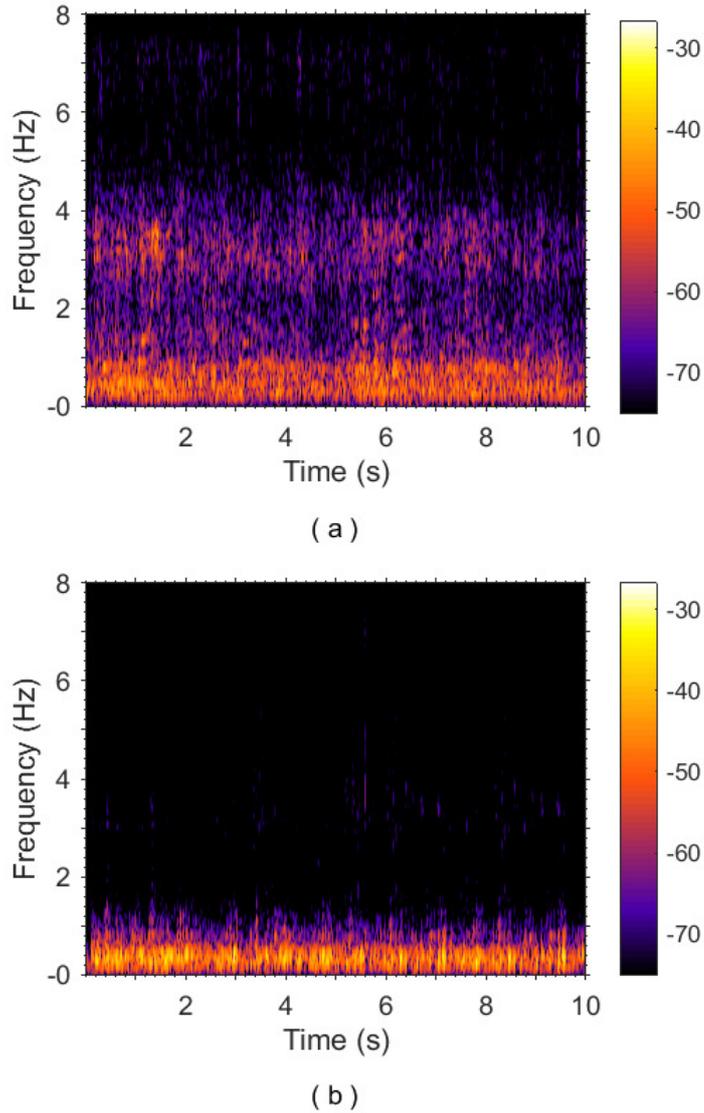

Fig. 2. Spectrogram of: (a) noise recorded inside a train wagon, and (b) cafeteria babble noise.

All signals were sampled at 16 kHz and were transformed to the frequency domain by a 512 point STFT using a 20 ms Hanning window with zero padding and 50% of overlap. After processed by each mask, the transformed signals in the STFT domain were reconstructed by the weighted overlap-and-add method [26]. The clean and noise signals were processed separately in order to calculate the ideal value of the masks on each frame.

The performance of each mask was assessed in terms of speech quality and intelligibility by using, respectively, the wideband Perceptual Evaluation of Speech Quality (PESQ) measure [27], and the Normalized Covariance Metric (NCM) [28]. Results were statistically compared through analysis of variance (ANOVA), with $p < 0.05$, followed by multiple comparison analysis using Tukey's test [29].

The noisy signals were divided in two groups: the training set, and the testing set. The training set was comprised by six subgroups, one for each SNR and type of noise, containing 100 speech files each. The testing set had also six subgroups, with a total of 620 speech files in each one.

The training set was applied for obtaining (by exhaustive search) the best sets of parameters for both CM and PW masks according to a proper objective quality criterion (PESQ, NCM, or a combination of both). Initially, the best parameter set was obtained for each one of the 600 training noisy signals according to an arbitrary grid of possibilities. For the CM, this grid was comprised by all combinations of $\mu$, varying from $-60$ dB to 60 dB in steps of 5 dB, and $\gamma$, varying from 0.5 to 1 as well as from 1.25 to 100, in 6 equally log-spaced steps, resulting in 300 different masks. The obtained sets of parameters include the WF ($\mu = \gamma = 1$) and a close approximation to the BM ($\mu = 1$, $\gamma = 100$). The same procedure was performed for the parametric Wiener filter, with $\beta$ varying from 0.2 to 1, and from 1.25 to 40 in, respectively, 6 and 4 logarithmically spaced steps, while $\eta$ was varied from $-35$ dB to 25 dB in 2.5 dB steps (totalizing 250 masks). After obtaining these 600 best sets, the median of the 100 results for each subgroup was calculated. This median set was then mapped to the nearest value in the arbitrary grid. The resulting mapped parameter sets, for each SNR and type of noise, were applied to the testing group.

### 4.1. Quality Maximization

In this first experiment, the method previously described for finding the best set of parameters, for each type of noise and SNR and for both CM and PW, was applied to maximize quality according to the PESQ criteria. The best sets of parameters obtained for each type of noise and SNR, using the training group, are shown in Table II. Table III shows the mean PESQ obtained for the testing group for each assessed mask, SNR, and type of noise.

Table II. Optimal parameter settings for the CM and PW mask. Maximization of the PESQ criteria in the training set.

| Noise | | Babble | | | Train | | |
|---|---|---|---|---|---|---|---|
| SNR [dB] | | −10 dB | 0 dB | 5 dB | −10 dB | 0 dB | 5 dB |
| CM | $\gamma$ | 0.66 | 0.66 | 0.66 | 0.57 | 0.5 | 0.57 |
| | $\mu$ [dB] | 10 | 5 | 5 | 25 | 20 | 15 |
| PW | $\beta$ | 0.72 | 0.53 | 0.53 | 0.53 | 0.53 | 0.53 |
| | $\eta$ [dB] | 7.5 | 5 | 5 | 15 | 12.5 | 10 |

Table III. PESQ results for WF, BM, CM, and PW, calculated from the testing set. Masks CM and PW were optimized for highest quality. For each column (same type of noise and SNR) the highest PESQ is bolded, while symbol § indicates results without statistical difference ($p < 0.05$).

| Noise | Babble | | | Train | | |
|---|---|---|---|---|---|---|
| SNR [dB] | −10 dB | 0 dB | 5 dB | −10 dB | 0 dB | 5 dB |
| WF | 1.364§ | 1.998 | 2.565 | 1.417 | 1.984 | 2.487 |
| BM | 1.141 | 1.578 | 2.068 | 1.178 | 1.557 | 2.041 |
| CM | **1.411** | **2.078**§ | **2.650**§ | **1.560**§ | **2.140**§ | **2.624** |
| PW | 1.375§ | 2.058§ | 2.629§ | 1.552§ | 2.115§ | 2.588 |

Fig. 3 shows bi-dimensional boxplots relating speech quality (PESQ) and intelligibility (NCM) scores for the unprocessed noisy-speech, speech processed by the WF, BM, CM, and PW for SNR = −10dB and babble noise. The horizontal axis presents the speech quality in PESQ units; while the vertical axis scores the intelligibility (unity corresponds to 100%). In this presentation form, the overall performance increases with the distance from the origin. Each square represents the limits of the first and third quartiles for the 620 sentences of the testing group. The outliers are omitted for clarity. The inset shows a zoom around the CM and PW medians, represented by plus signs (+). For other SNRs and the inside train noise, the boxplots are similar and are not presented.

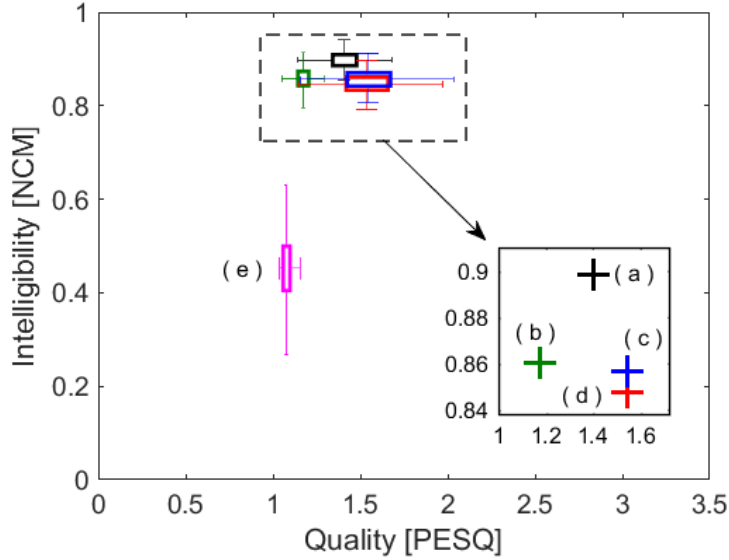

Fig. 3. Intelligibility and quality bi-dimensional boxplots for the testing set and inside train noise at SNR = −10 dB. Parameters for the CM and PW were optimized for quality. (a) Wiener (black); (b) BM (green); (c) CM (blue); (d) PW (red); (e) noisy (magenta).

Table IV. Optimal parameter settings for the CM and PW mask. Maximization of the NCM criteria in the training set.

| Noise | | babble | | | train | | |
|---|---|---|---|---|---|---|---|
| SNR [dB] | | −10 dB | 0 dB | 5 dB | −10 dB | 0 dB | 5 dB |
| CM | $\gamma$ | 0.66 | 0.76 | 0.76 | 0.76 | 0.87 | 0.87 |
| | $\mu$ [dB] | −5 | −5 | −5 | −5 | −10 | −10 |
| WP | $\beta$ | 0.53 | 0.53 | 0.72 | 0.53 | 0.72 | 1.25 |
| | $\eta$ [dB] | 0 | 0 | −2.5 | 2.5 | −2.5 | −5 |

### 4.2. Intelligibility Maximization

In this second experiment, the same procedure for finding the best set of parameters were applied to the CM and PW, but with the aim of maximizing intelligibility according to the NCM criteria. Table IV shows the best sets of parameters obtained for each type of noise and SNR (using the training group), while Table V presents the mean NCM obtained for the testing group (for each assessed mask, SNR and type of noise). Fig. 4 shows bi-dimensional boxplots relating speech quality (PESQ) and intelligibility (NCM) scores for the unprocessed noisy-speech, speech processed by the WF, BM, CM, and PW, for SNR = −10dB and inside train noise. The inset shows a zoom around the CM and PW medians (plus signs). Boxplots for other SNRs and babble noise are similar and not presented.

Table V. NCM values for WF, BM, CM, and PW calculated from the testing set. Masks CM and PW were optimized for highest intelligibility. For each column (same type of noise and SNR) the highest NCM is bolded, while symbol § indicates results without statistical difference ($p < 0.05$).

| Noise | babble | | | train | | |
|---|---|---|---|---|---|---|
| SNR [dB] | −10 dB | 0 dB | 5 dB | −10 dB | 0 dB | 5 dB |
| WF | 0.832 | 0.948 | 0.984 | 0.897 | 0.963 | 0.989 |
| BM | 0.733 | 0.908 | 0.967 | 0.858 | 0.950 | 0.984 |
| CM | **0.881**$^§$ | **0.967**$^§$ | **0.992**$^§$ | **0.911**$^§$ | **0.971**$^§$ | **0.992**$^§$ |
| PW | 0.879$^§$ | 0.966$^§$ | 0.992$^§$ | 0.910$^§$ | 0.971$^§$ | 0.992$^§$ |

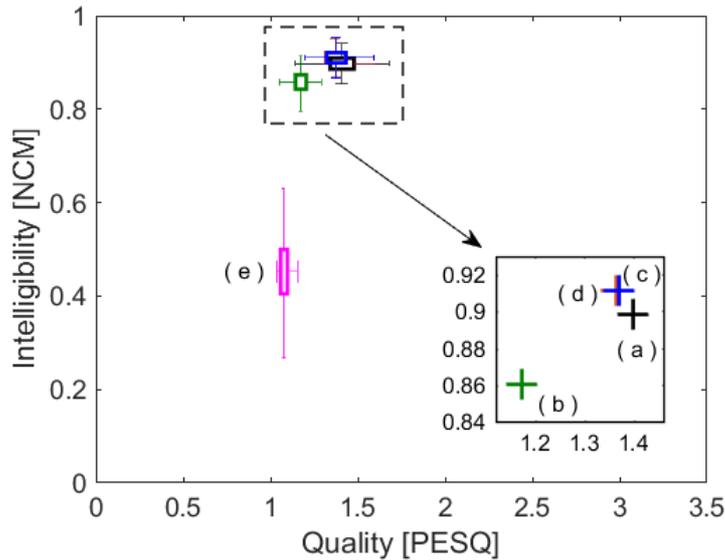

Fig. 4. Intelligibility and quality bi-dimensional boxplots for the testing set and inside train noise at SNR = −10 dB. Parameters of the CM and PW were optimized for intelligibility. (a) Wiener (black); (b) BM (green); (c) CM (blue); (d) PW (red); (e) noisy (magenta).

### 4.3. Quality and Intelligibility Maximization

In this experiment, the CM and PW parameters were optimized for providing the best overall performance in terms of both quality and intelligibility. The overall performance was calculated as a quadratic distance given by:

$$d_p = \text{NCM}^2 + \left(\frac{\text{PESQ}}{5}\right)^2, \qquad (25)$$

since $0 \leq \text{NCM} \leq 1$ and $0 \leq \text{PESQ} \leq 5$. The optimum parameter sets that lead to maximization of (25) are shown in Table VI. Tables VII and VIII show NCM and PESQ results for, respectively, babble and inside train noise. Fig. 5 shows the boxplot for the joint optimization of both intelligibility and quality measures.

Table VI. Optimal parameter settings for the CM and PW mask. Maximization of (25) in the training set.

| Noise | | babble | | | train | | |
|---|---|---|---|---|---|---|---|
| SNR [dB] | | −10 dB | 0 dB | 5 dB | −10 dB | 0 dB | 5 dB |
| CM | $\gamma$ | 0.66 | 0.66 | 0.76 | 0.66 | 0.66 | 0.66 |
| | $\mu$ [dB] | −5 | −5 | 0 | 0 | 0 | 0 |
| WP | $\beta$ | 0.53 | 0.72 | 0.72 | 0.53 | 0.53 | 0.53 |
| | $\eta$ [dB] | 0 | 0 | 0 | 0 | 5 | 5 |

Table VII. NCM and PESQ for maximizing (25) and babble noise. For each column (same type of noise and SNR) the highest NCM and PESQ are bolded, while symbol § indicates results without statistical difference ($p < 0.05$).

| | NCM | | | PESQ | | |
|---|---|---|---|---|---|---|
| | −10 dB | 0 dB | 5 dB | −10 dB | 0 dB | 5 dB |
| WF | 0.832 | 0.948 | 0.984 | 1.364[§] | 1.998[§] | 2.565 |
| BM | 0.733 | 0.908 | 0.967 | 1.141 | 1.578 | 2.068 |
| CM | **0.881** | **0.967** | **0.989** | 1.376[§] | 1.999[§] | **2.633**[§] |
| PW | 0.867 | 0.955 | 0.987 | **1.400** | **2.043** | 2.611[§] |

Table VIII. NCM and PESQ for maximizing (25) and inside noise train. For each column (same type of noise and SNR) the highest NCM and PESQ are bolded, while symbols § and * indicate pair of results without statistical difference ($p < 0.05$).

|     | NCM | | | PESQ | | |
| --- | --- | --- | --- | --- | --- | --- |
|     | −10 dB | 0 dB | 5 dB | −10 dB | 0 dB | 5 dB |
| WF  | 0.897 | 0.963 | **0.989**§ | **1.417**§ | 1.984§ | 2.487§ |
| BM  | 0.858 | 0.950 | 0.984* | 1.178 | 1.557 | 2.041 |
| CM  | **0.910**§ | **0.968** | **0.990**§ | 1.409§ | 2.010§ | 2.518§ |
| WP  | 0.908§ | 0.959 | 0.984* | **1.427**§ | **2.068** | **2.564** |

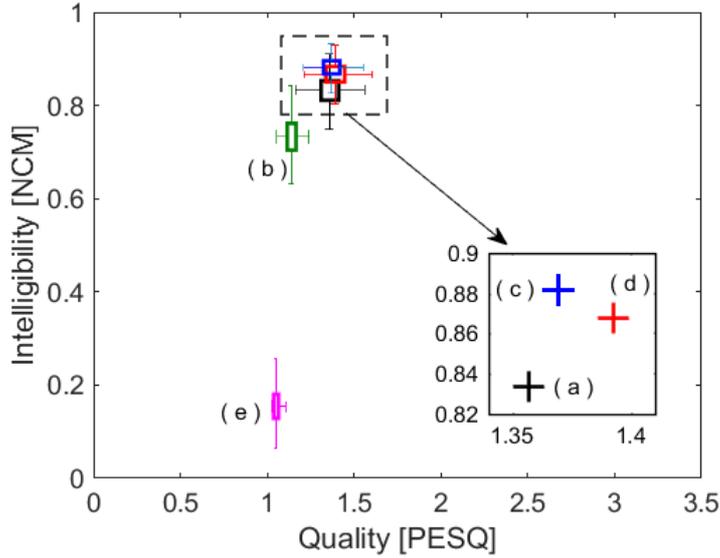

Fig. 5. Intelligibility and quality bi-dimensional boxplots for the testing set and babble noise at SNR = −10dB. Parameters of the CM and PW were optimized for maximizing (25). (a) Wiener (black); (b) BM (green); (c) CM (blue); (d) PW (red); (e) noisy (magenta).

## 5. DISCUSSION

The proposed noise suppression mask, equation (23), was obtained from a meaningful cost function that establishes a trade-off between noise reduction and speech distortion. Its convexity was demonstrated and thus proved that there is a global minimum. Its parameters provide two degrees of freedom. The parameter $\mu$ is associated to the decision threshold, while $\gamma$ controls the slope with relation to the local SNR. Differently from the widely-used parametric Wiener filter presented in (6), in which those both characteristics are simultaneously affected by $\beta$ (see Fig. 6), each parameter of the proposed mask independently controls a given feature. This property is especially interesting for those applications that support online adaptation of the suppression rule due to variations of the acoustic scenario [30]. This is the case of speech enhancement systems for real-time embedded applications such as hearing aids and cochlear implants.

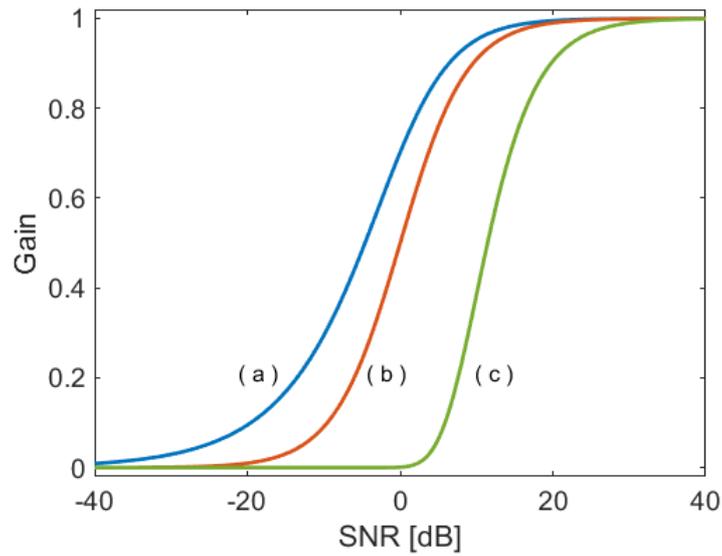

Fig. 6. Suppression function of the parametric Wiener filter (equation (6)) for $\eta = 0$ dB and: (a) $\beta = 0.51$ (blue); (b) $\beta = 1$ (red); (c) $\beta = 10$ (green).

Another important finding with relation to the conformability of the proposed mask is that, despite the apparent similarity between both generation functions (equations (6) and (23)), the CM and the PW filter do not match perfectly. In fact, CM may provide accurate approximations to the PW for a large range of pairs $(\beta,\eta)$, while the inverse is not true.

Fig. 7 exemplifies the conformability limits of PW for approximating the CM in two different situations: (a) CM as a soft-mask ($\gamma < 1$); and (b) CM as a hard-mask ($\gamma \gg 1$). The PW parameter pairs $(\beta,\eta)$ were optimized by minimization of the RMSE criteria, defined in (24), using the Nelder-Mead simplex algorithm [31], for obtaining the best approximation to a given setting of the CM parameters $(\gamma, \mu)$[2]. The RMSE for the curves depicted on Fig. 7a and Fig. 7b are, respectively, 0.79 and 1.07. Fig. 7b indicates that the PW filter cannot precisely approximate the morphology of the CM for very large slopes ($\gamma \gg 1$). On the other hand, additional simulations showed that CM provides accurate values for approximating PW, resulting in a maximum RMSE of 0.14 for all studied situations.

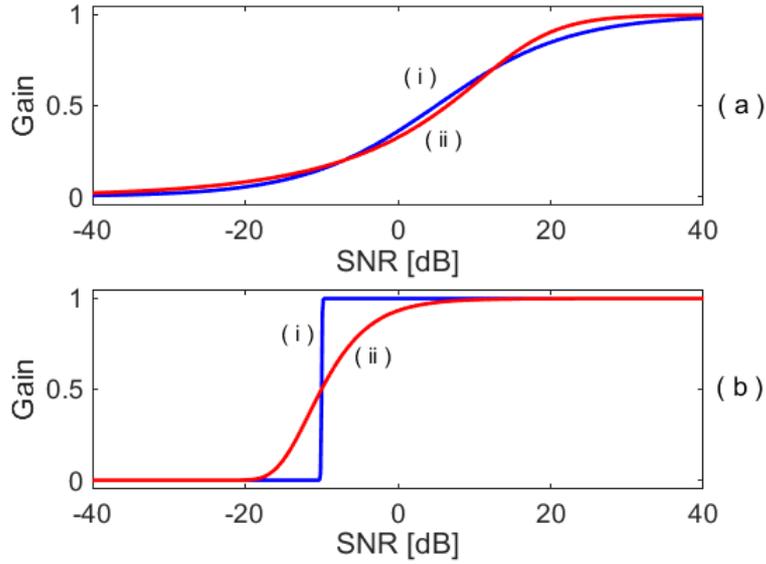

Fig. 7. Approximation of PW (red) to CM (blue). (a) $\eta = 15.9$ dB, $\beta = 0.305$, $\mu = 5$ dB, and $\gamma = 0.5$; (b) $\eta = -42.6$ dB, $\beta = 1.25 \times 10^3$, $\mu = -10$ dB, $\gamma = 100$. Parameters of PW were chosen to minimize the RMSE with respect to CM.

When maximized for quality (Section 4.1), CM overcomes or, at least, achieves the same performance of all assessed techniques (WF, BM, and PW). For maximizing intelligibility (Section 4.2) both CM and PW show the same performance ($p < 0.05$), overcoming results obtained by both WF and BM. By changing the parameter settings of both CM and PW, it is possible to change the trade-off between quality and intelligibility performance. This demonstrates the versatility of these masks and their potential for adjusting to the individual preferences of the listener.

---

[2] In this experiment the values of $(\beta, \eta)$ are not constrained to the arbitrary grid defined in Section 4.

In all experiments the BM performance was significantly worse as compared to WF, CM and PW soft masks. This corroborates the experimental results reported by [3] [7].

In general, for both types of noise and SNRs, CM has the best intelligibility performance, while the PW leads to the best quality (PESQ) results. Nevertheless, these two masks perform perceptually in a similar way, considering the minimum noticeable PESQ difference is around 0.2 [32] [33].

Simulation results showed that, for all optimal (quality and/or intelligibility) cases analyzed, the value of parameter $\gamma$ is smaller than 1, which corresponds to softer masks when compared to the Wiener filter. It can also be observed that for increased quality $\mu$ should be greater or equal to 0 dB, whilst for increased intelligibility $\mu < 0$ dB.

Despite a general preference for soft masks by normal hearing subjects, the Binary mask is also of significant relevance, as experiments with hearing-impaired listeners revealed great variability of inter-subject performance [6] [7], and for certain scenarios there may be individual preferences in favor of a hard mask approach. Thus, a more versatile mask such as the CM may be desirable for hearing aid and cochlear implant users.

Considering $\gamma \to \infty$ (see Table I), thus $\alpha \to \frac{1}{2}_+$, and the proposed cost function presented in (11) can be interpreted as the weighted sum of the square root of both speech and noise distortions. This observation supports the theoretical optimality of the Binary mask under such context. The optimality of the BM has been addressed before. Arguments for supporting global optimality of the ideal binary mask ($\mu_0 = 1$), as compared to all binary masks, were provided in [34]. In [35], it was demonstrated that, assuming that the magnitude-squared spectrum of the noisy speech signal can be approximated by the sum of the clean signal and noise magnitude-squared spectrum, the BM is the Maximum a Posteriori estimator of the magnitude-squared spectrum. Nevertheless, a theoretical proof of the optimality of the BM, in the context of a cost function minimization problem, had never been demonstrated before.

# 6. CONCLUSION

This work shows that a previously heuristically designed version of the parametric Wiener filter is, in fact, the optimal time-frequency mask that minimizes a trade-off between noise and speech distortions. The associated cost-function is defined by a weighted sum of powers of speech and noise magnitude distortions.

The resulting mask allows, by the setting of its parameters, the independent adjustment of the slope and bias of the suppression function with respect to the local SNR. This may be a desired characteristic for applications that require online adaptation of the suppression function according to variations of the acoustic scenario. This is the case of speech enhancement systems for real-time applications such as hearing aids and cochlear implants

Simulation results indicate that the proposed time-frequency mask, called Conformable Wiener filter, has similar psychoacoustic performance as compared to the well-known heuristically-designed Parametric Wiener filter in terms of quality (PESQ). For specific situations, it may present higher intelligibility as compared to the parametric Wiener filter.

A relevant characteristic of the Conformable mask is that it can approximate time-frequency hard-masks with higher accuracy than the classical parametric Wiener. This characteristic has shown to be important in experiments with hearing impaired listeners in certain acoustic scenarios.

As a result, this work provided a unified framework for deriving and interpreting the optimality of a family of time-frequency masks that encompasses the well-known Wiener filter and the Binary mask, as well as some of their variations.

# ACKNOWLEDGMENTS

This work was supported by the Brazilian Ministry of Science and Technology (CNPq) under grant 304867/2015-2.